\documentclass[12pt]{article}
\usepackage{graphicx}
\usepackage{amsmath}

\newcommand\pubnumber{SNSN-323-63}
\newcommand\pubdate{\today}

\def\cea{CEA Saclay\\
91000 Gif-sur-Yvette, FRANCE}
\def\mainz{Institut fuer Kernphysik\\
Johannes-Gutenberg Universitaet Mainz, 55128 Mainz, GERMANY}

\def\Title#1{\begin{center} {\Large #1 } \end{center}}
\def\Author#1{\begin{center}{ \sc #1} \end{center}}
\def\Address#1{\begin{center}{ \it #1} \end{center}}

\newcommand\pubblock{\rightline{\begin{tabular}{l} \pubnumber\\
         \pubdate  \end{tabular}}}
\newenvironment{Abstract}{\begin{quotation}  }{\end{quotation}}
\newenvironment{Presented}{\begin{quotation} \begin{center}
             PRESENTED AT\end{center}\bigskip
      \begin{center}\begin{large}}{\end{large}\end{center} \end{quotation}}

\def\beq{\begin{equation}}
\def\eeq#1{\label{#1}\end{equation}}
\def\eeqn{\end{equation}}

\def\beqa{\begin{eqnarray}}
\def\eeqa#1{\label{#1}\end{eqnarray}}
\def\eeqan{\end{eqnarray}}

\let\bar=\overbar

\def\Dslash{\not{\hbox{\kern-4pt $D$}}}
\def\dslash{\not{\hbox{\kern-2pt $\del$}}}

\def\msb{{\bar{\ssstyle M \kern -1pt S}}}

\begin{document}
\begin{titlepage}
\pubblock

\vfill
\Title{Hadron Multiplicity and Fragmentation in SIDIS}
\vfill
\Author{ Nicolas Pierre}
\Address{\cea \\ \mainz}
\vfill
\begin{Abstract}
COMPASS final results on multiplicities of charged hadrons and of identified pions and kaons produced in the deep inelastic muon scattering off
an isoscalar target are presented and compared to HERMES results. Measurements are done in bins of x, y and z in a wide kinematic range.
The hadron and pion data show a good agreement with (N)LO QCD expectations. The kaon data are long awaited for since they are needed to
extract kaon fragmentation functions, a crucial ingredient in solving the strange quark polarisation puzzle. COMPASS results for kaons differ
from the expectations of the old NLO DSS fit and they cannot be well described by LO QCD either. In this context the importance of
$K^-$/$K^+ˆ'$ multiplicity ratio at high z is discussed.
\end{Abstract}
\vfill
\begin{Presented}
CIPANP 2018 - Thirteenth Conference on the Intersection of Particle and Nuclear Physics\\
Indian Wells, United States of America,  May 29 -- June 3, 2018
\end{Presented}
\vfill
\end{titlepage}
\def\thefootnote{\fnsymbol{footnote}}
\setcounter{footnote}{0}

\section{Introduction : Quark Fragmentation Functions}

Quark Fragmentation Functions are non-perturbative objects that are needed to describe several processes.
While light flavoured Quark Fragmentation Functions like up quark and down quark are well determined and
constrained, the strange quark Fragmentation function is still not well known. Moreover the largest
uncertainty in the $\Delta s$ extraction from polarized SIDIS data comes from the bad knowledge of strange
quark fragmentation function into kaons. The data available from $e^+e^-$ and $pp$ collisions are not sufficient
and at too high $Q^2$ to allow a good determination of the strange quark Fragmentation Function.

In order to access the quark fragmentation functions in SIDIS, one measures the multiplicity of hadrons
$lN \rightarrow l'hX$ where $h=\pi,K,p$.
In the expression of the multiplicities, the Parton Distribution Functions (PDFs) and the quark Fragmentation Functions (FFs) are linked :
\begin{equation}
  \frac{dM^h(x,y,z)}{dz} \overset{LO}{=} \frac{\sum_q e^2_q q(x,Q^2) D^h_q(z,Q^2)}{\sum_q e^2_q q(x,Q^2)},\,\, z=\frac{E_h}{\nu}=\frac{E_h}{E_{\nu}-E_{\nu'}}
\end{equation}

One can access the quark FF assuming the PDF known. It has to be noted that while the PDFs depend on $x$, the FFs depend on $z$.
When the kaon multiplicities are measured, one can constrain the product $s(x,Q^2).D^K_s(z,Q^2)$.

Thus, pion and kaon multiplicities allow to do global NLO QCD analyses to extract quark FFs and the strangeness contained in the kaons.

\section{Multiplicity data}

\begin{figure}[!htb]
\centering
\includegraphics[height=7.3cm]{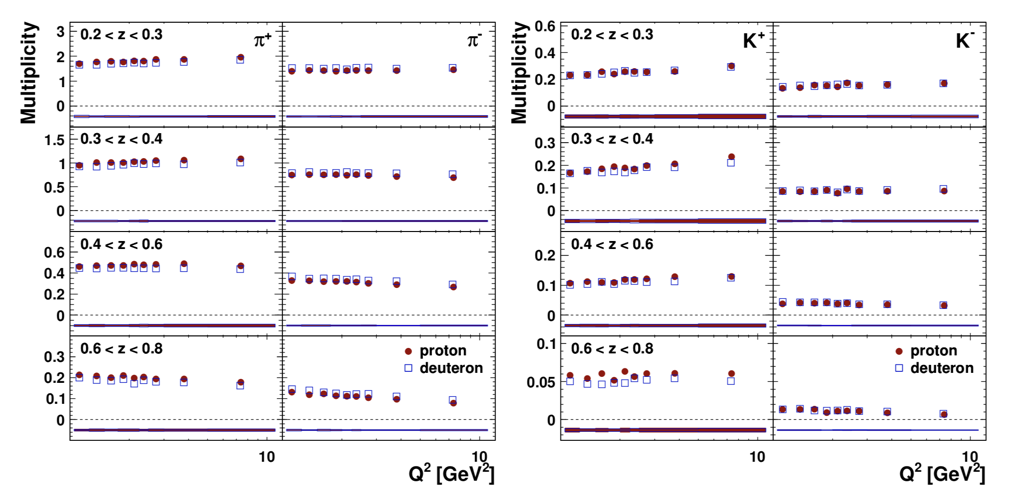}
\caption{From left to right, HERMES multiplicities of $\pi^+$, $\pi^-$, $K^+$, $K^-$ versus $Q^2$ in a range
going from 1 to 15 GeV$^2$ in bins of $z$. Proton data are in red dots, deuteron data in blue squares.}
\label{fig:HERMES}
\end{figure}

\begin{figure}[!htb]
\centering
\includegraphics[height=8cm]{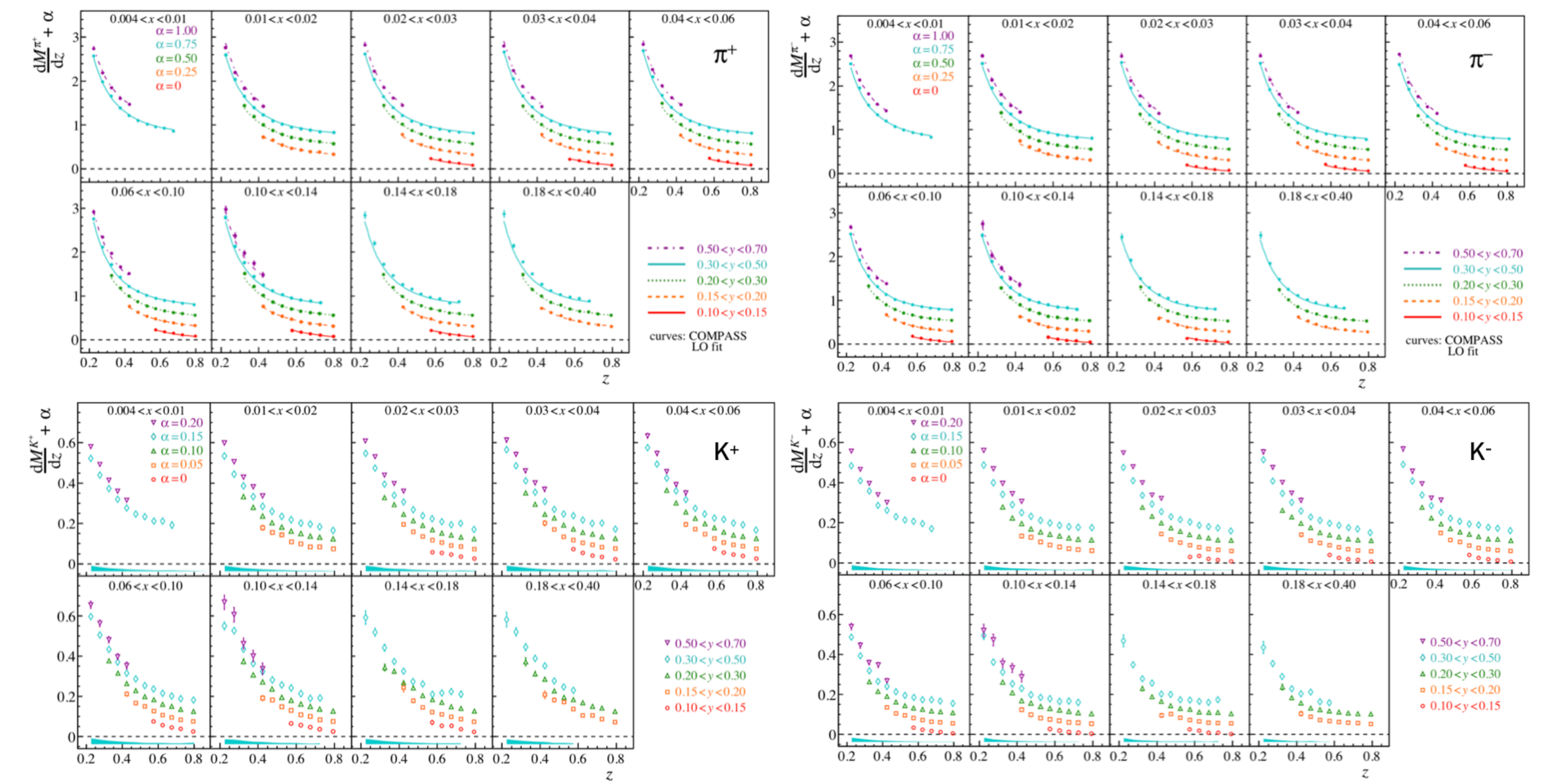}
\caption{From left to right and top to bottom, COMPASS multiplicities of $\pi^+$, $\pi^-$, $K^+$, $K^-$ versus z in bins of $x$
and staggered vertically with $y$ with different colors. The entire set comprise more than 1200 data points in total with a strong
$z$ dependence. One can note that $M^{\pi^{+}}\sim M^{\pi^{-}}$ and $M^{K^{+}} > M^{K^{-}}$}
\label{fig:COMPASS}
\end{figure}

Two experiments have released pion and kaon multiplicity from SIDIS : HERMES\cite{HERMES} and COMPASS\cite{COMPASS}.
HERMES released pion and kaon multiplicity data on both proton and deuteron target while COMPASS released
pion and kaon multiplicity data only on deuteron target (Fig.\ref{fig:HERMES} and \ref{fig:COMPASS}).

\section{Extraction of Quark Fragmentation Functions from hadron Multiplicities}

\begin{figure}[!htb]
\centering
\includegraphics[height=4.8cm]{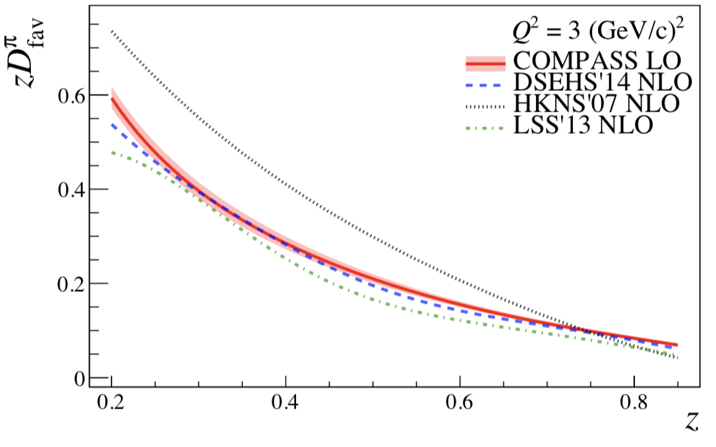}\includegraphics[height=4.8cm]{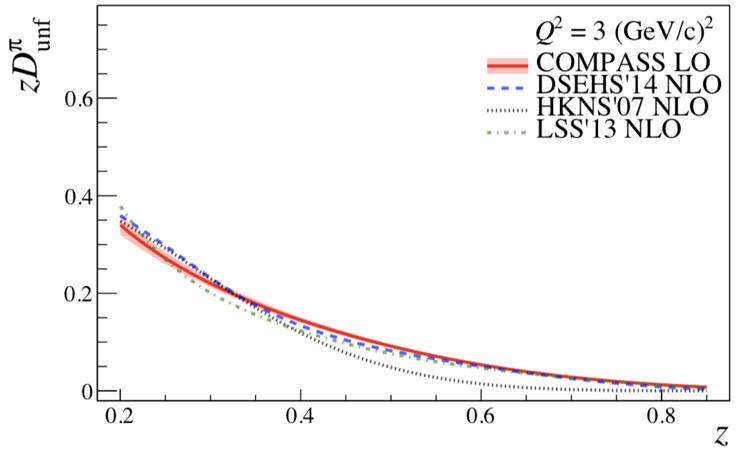}
\caption{COMPASS LO fits assuming two independent fragmentation functions $D^{\pi}_{fav}$ and $D^{\pi}_{unf}$.
As expected, $D^{\pi}_{fav} > D^{\pi}_{unf}$ and COMPASS LO fit agrees with the fits of DSEHS NLO and LSS NLO.}
\label{fig:Frag}
\end{figure}

COMPASS has performed a Leading Order (LO) fits of pion and kaon multiplicities in order
to extract quark fragmentation from it. In the pion case, the number of independent FFs
can be reduced to 2 : $D^{\pi}_{fav}$ and $D^{\pi}_{unf}$. In Fig.\ref{fig:Frag}, $D^{\pi}_{fav}$
is above $D^{\pi}_{unf}$ as expected. It has also to be noted that COMPASS LO agrees with DSEHS and LSS LO.

In the kaon case, the number of independent FFs can be reduced to 3 : $D^{K}_{fav}$,
$D^{K}_{str}$ and $D^{K}_{unf}$. The LO fit is not conclusive due to difficulties fitting
high $z$ data even at NLO. These difficulties will be discussed later.
The DSS17 fit includes COMPASS kaon data and finds a smaller $D^{K}_{str}$ than
previously obtained (DSS07 was largely based on HERMES old data). This is confirmed by a new
iterative study of strange PDF and FF with BSS\cite{BSS} in order to bring more constrains to the fit.
An other way to constrain FFs is to look at the $K^++K^-$ multiplicity sum.

\section{Sum of z-Integrated Multiplicities}

For an isoscalar deuteron target, the sum of $\pi^+$ and $\pi^-$ multiplicities integrated over $z$ reads at LO :
\begin{equation}
  \begin{split}
    M^{\pi^++\pi^-}(x) &= D^{\pi}_{fav} +D^{\pi}_{unf} - \frac{2(s(x)+\bar{s}(x))(D^{\pi}_{fav} - D^{\pi}_{unf})}{5(u(x)+d(x)+\bar{u}(x)+\bar{d}(x))+2(s(x)+\bar{s}(x))} \\
    &\overset{high\,x}{=} D^{\pi}_{fav} +D^{\pi}_{unf}
  \end{split}
\label{eq:sumpion}
\end{equation}

\begin{figure}[!htb]
\centering
\includegraphics[height=5cm]{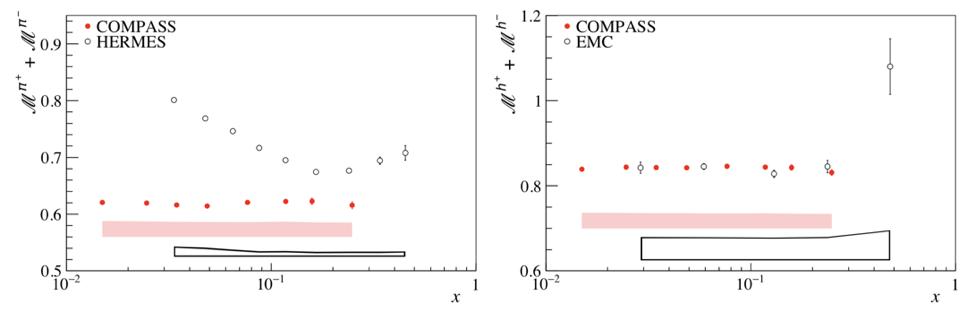}
\caption{From left to right, multiplicity sum of pions and hadrons. COMPASS pion sum is significantly below HERMES ones. One can also
note that there is no x dependence for COMPASS pion sum as in EMC hadrons sum.}
\label{fig:SumPion}
\end{figure}

For this quantity, $M^{\pi^+ + \pi^-}$ COMPASS data are significantly below HERMES ones (Fig.\ref{fig:SumPion} left). The absence of x dependence
for COMPASS pion data follows the expectation of Eq.\ref{eq:sumpion}. As hadrons are mostly pions, the results
of the multiplicity sum for hadrons should be comparable. EMC have done the
measurement of hadrons multiplicities. They concur with measurement of COMPASS, namely
there is no x dependence in hadron multiplicity sum hence in pion multiplicity sum (Fig.\ref{fig:SumPion} right).

For an isoscalar deuteron target, the sum of $K^+$ and $K^-$ multiplicities integrated over $z$ reads at LO :
\begin{equation}
  \begin{split}
    M^{K^++K^-}(x) &= \frac{(u+d+\bar{u}+\bar{d})(4D^{K}_{fav}+6D^{K}_{unf})+(s+\bar{s})(D^{K}_{str}+D^{K}_{unf})}{5(u(x)+d(x)+\bar{u}(x)+\bar{d}(x))+2(s(x)+\bar{s}(x))} \\
    &= \frac{QD^{K}_{Q}+SD^{K}_{S}}{5Q+2S} \overset{high\,x,\,S\sim0}{=} \frac{D^{K}_{Q}}{5}
  \end{split}
\end{equation}

\begin{figure}[!htb]
\centering
\includegraphics[height=6cm]{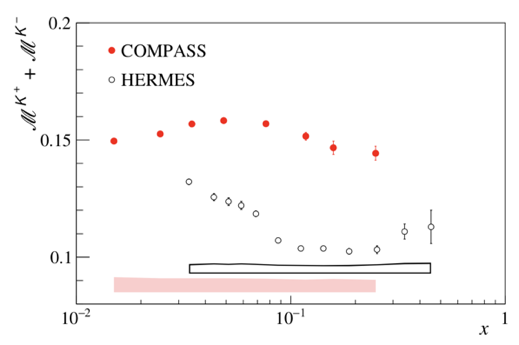}
\caption{Multiplicity sum of kaons. COMPASS kaon sum are significantly above HERMES results. One can
note that the multiplicity sum is steady at low x, thus pointing to a smaller $D^K_S$ than previous NLO fits.}
\label{fig:SumKaon}
\end{figure}

COMPASS kaon data are significantly above HERMES ones (Fig.\ref{fig:SumKaon}).
At high x, one can assume that $s(x)$ is close to zero. COMPASS data point to a higher $D^{K}_{Q}$
(paper value < $D^{K}_{Q}$ < paper value) than previous NLO fits did ($D^{K}_{Q} \sim 0.43 \pm 0.04$).
At low x, one can assume that $D^{K}_{str} > D^{K}_{fav}$ because the sea contribution is larger than
the valence one.

\subsection*{Comments on radiative corrections}

In the COMPASS paper of kaon multiplicities, TERAD was used for the inclusive corrections
viz. the corrections of the lepton line. TERAD without the elastic and quasi-elastic tails
was used for the semi-inclusive corrections viz. the corrections of the hadron line.
Since TERAD was giving z-integrated semi-inclusive corrections it was used to set an
upper limit to the correction to apply, which was between $1\%$ and
$7\%$ depending on kinematics.
COMPASS worked on a way to have a z-dependent semi-inclusive correction and is now using
DJANGOH, which accounts for this z-dependence. Comparison between DJANGOH and TERAD were
conducted. For inclusive corrections, an agreement between DJANGOH and TERAD was reached
within $3\%$, which is a really good achievment taking into account the fact that the two
frameworks use a different renormalization scheme. Concerning the semi-inclusive corrections
for kaons, the correction is between $0\%$ and $10\%$ depending on the kinematics with an
average at $5\%$. DJANGOH corrections fall well within the systematic errors in the publication.

\section{Ratio of z-integrated kaon multiplicities $M^{K^-}/M^{K^+}$}

The kaon multiplicity ratio $M^{K^-}/M^{K^+}$ is an interesting quantity to observe as several experimental
and theoretical uncertainties cancel (acceptance, radiative correction etc.).
%
%
%

\begin{figure}[!htb]
\centering
\includegraphics[height=7cm]{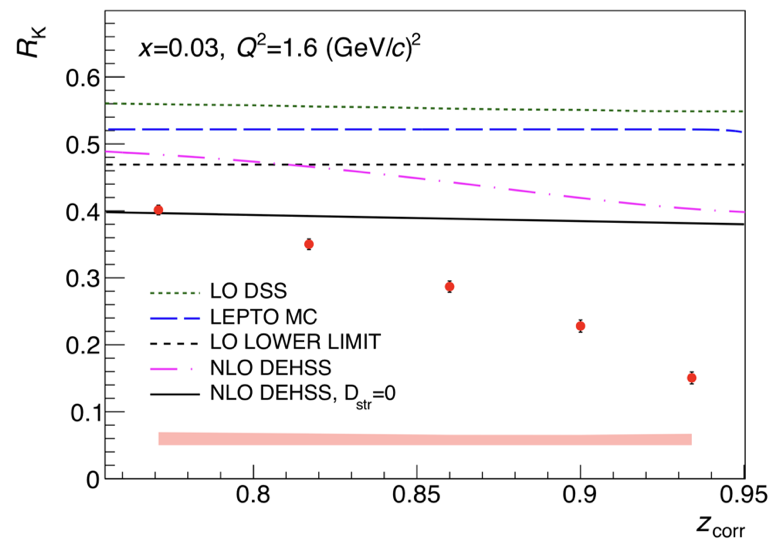}
\caption{$R_K = K^-/K^+$ ratio at LO for COMPASS data compared to several models and LO/NLO calculations.}
\label{fig:RK}
\end{figure}

However, when comparing the obtained $R_K$ with predictions\cite{KaonRatio}, there is a clear disagreement
with the models and with the LO/NLO calculations (Fig.\ref{fig:RK}).

In order to understand this discrepancy, one can look at the dependence $R_K$ can have
with other kinematic variables. For example, a clear $\nu$ dependence of $R_K$ can be
observed : as $\nu$ increases, $R_K$ gets closer to pQCD expectations.

Going further in this direction, considering now high-z kaons, the phase space is reduced
for other particles. As there are conservation laws, the missing mass $M_X$ is useful to
study this region :

\begin{equation}
  M_X \approx \sqrt{M^2_p+2M_p\nu(1-z)-Q^2(1-z)^2}
\end{equation}

\begin{figure}[!htb]
\centering
\includegraphics[height=7cm]{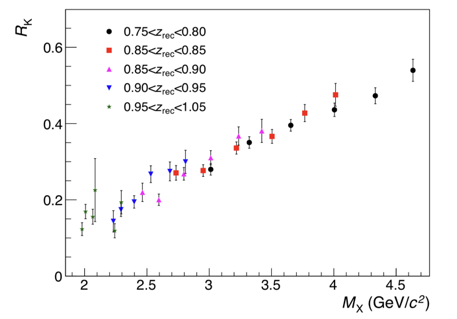}
\caption{Correlation plot between the missing mass $M_X$ and the kaon multiplicity ratio $R_K$
for different z bins. The two quantities look to be highly correlated.}
\label{fig:MX}
\end{figure}

The Fig.\ref{fig:MX} shows that $R_K$ and $M_X$ are highly correlated while a flat value of $R_K$ is expected
from pQCD calculations. It suggests that a correction within the pQCD formalism is needed to take into
account the available phase space for target remnant hadronization.

HERMES and COMPASS results show the same trend in the kaon multiplicity ratio nevertheless
COMPASS data are below HERMES ones. However as seen previously, there is a strong $\nu$ dependence
of $R_K$ and this may contribute to observed disagreement. A direct comparison with the only common
kinematic data point between COMPASS and HERMES data can be made and there an agreement is reached.
In the neighbouring x bins, HERMES $<\nu>$ is smaller than COMPASS one and HERMES ratio is smaller
than COMPASS one, again pointing to the same kind of dependence described supra.
Note also that target mass correction are suggested by J.V. Guerrero and A. Accardi\cite{Accardi} which could explain part of
the discrepancy in kaon data.

\section{Conclusion and Open Questions}

HERMES and COMPASS offer samples of pion and kaon multiplicities measured in SIDIS. COMPASS
offers the largest sample of kaon multiplicities (over 600 data points) to constrain Fragmentation Functions.
The z-integrated multiplicity sum shows large discrepancies between COMPASS and HERMES for both
pion and kaon multiplicity sums, going up to $25\%$ in the z-integrated $M^{\pi^+}+M^{\pi^-}$
with COMPASS below HERMES and up to $30-40\%$ in the z-integrated $M^{K^+}+M^{K^-}$ with
COMPASS above HERMES.
The kaon multiplicity ratio $R_K = M^{K^-}/M^{K^+}$, measured for the first time at high $z$ at COMPASS,
shows a large disagreement with pQCD calculations at high $z$ and low $\nu$.
The correlation of $M_X$ and $R_K$ suggests that a correction within pQCD formalism is needed to take into
account the available phase space for the target remnant hadronization.

\end{document}